\documentclass{article}
\usepackage[T1]{fontenc}
\usepackage{a4wide}
\makeatletter

\newcommand{\LyX}{L\kern-.1667em\lower.25em\hbox{Y}\kern-.125emX\@}

\newcommand{\lyxaddress}[1]{
  \par {\raggedright #1 
  \vspace{1.4em}
  \noindent\par}
}

\makeatother

\begin{document}

\title{Teleporting Noncommuting Qubits Require Maximal Entanglement}

\author{Sibasish Ghosh\protect\( ^{1}\protect \)\thanks{
res9603@isical.ac.in
}, Guruprasad Kar\protect\( ^{1}\protect \), Anirban Roy\protect\( ^{1}\protect \)\thanks{
res9708@isical.ac.in
}, Debasis Sarkar\protect\( ^{2}\protect \) and Ujjwal Sen\protect\( ^{3}\protect \)\thanks{
dhom@boseinst.ernet.in
}}

\maketitle

\lyxaddress{\protect\( ^{1}\protect \)Physics and Applied Mathematics Unit, Indian Statistical
Institute, 203 B. T. Road, Calcutta 700035, India }

\lyxaddress{\protect\( ^{2}\protect \)Department of Mathematics, Burdwan University, Burdwan,
West Bengal, India }

\lyxaddress{\protect\( ^{3}\protect \)Department of Physics, Bose Institute, 93/1 A. P.
C. Road, Calcutta 700009, India}

\begin{abstract}
\noindent{Very recently, it was shown by Ghosh, Kar, Roy and Sen (\emph{Entanglement
vs. Noncommutativity in Teleportation}, quant-ph/0010012) that if it is \emph{a
priori} known that the state to be teleported is from a commuting set of qubits,
a separable channel is sufficient. We show that 1 ebit of entanglement is a
necessary resource to teleport a qubit even when it is known to be one of two
noncommuting states.}
\end{abstract}
Quantum teleportation is a process by which an exact replica of an unknown qubit
surfaces, without being physically transported, at a possibly distant location
while all information about the qubit vanishes at its original location. The
last phrase is a necessity by virtue of the no-cloning theorem\cite{1}.

A protocol realising this was proposed by Bennett \emph{et al.}(BBCJPW)\cite{2}
which required a maximally entangled channel between the sender say, Alice and
the receiver say, Bob. It has been shown in refs.\cite{3,4}, that teleportation
of an \emph{unknown} \( d \)-dimensional state (not necessarily a qubit) requires
maximal entanglement in \( d\otimes d \) between the sender and the receiver
(See also \cite{5}). Henceforth we consider only qubits. 

But what if some \emph{a priori} information is available about the qubit which
is to be teleported? Would it then be possible to teleport that qubit through
a non-maximally entangled channel? Ghosh \emph{et al.}\cite{5} have shown that
even to teleport any set of two noncommuting qubits, entanglement of the channel
is \emph{necessary}. Here we show that if the state to be teleported is known
to be one of two noncommuting qubits, the channel must be \emph{maximally} entangled
in \( 2\otimes 2 \), i.e., the sender and receiver must share 1 ebit. In ref.\cite{5}
it was shown that a separable channel is \emph{sufficient} to teleport qubits
(generalized to qu\emph{dits} in ref. \cite{6}) which are \emph{a priori} known
to be commuting. 

First of all we show that even if the qubit to be teleported is known to be
one of two non-orthogonal \emph{pure} qubits, the channel must be maximally
entangled in \( 2\otimes 2 \). Suppose that a source delivers a particle (call
it particle 1) in the state \( \left| \chi _{1}\right\rangle  \) or \( \left| \chi _{2}\right\rangle  \)
to Alice, where \( \left| \chi _{1}\right\rangle  \) and \( \left| \chi _{2}\right\rangle  \)
are arbitrary but fixed non-orthogonal qubits. Alice has the task of teleporting
the delivered qubit to a possibly distant party Bob, by using only local operations
and classical communication and any shared state \( \rho ^{AB2}_{ch} \) as
channel. Here A is with Alice whereas B and 2 are with Bob, 2 being the particle
at which the teleported state is to surface. 

Whatever local protocol Alice and Bob adheres to, it can be represented as 
\begin{equation}
\label{1}
(\left| \chi _{j}\right\rangle \left\langle \chi _{j}\right| )^{1}\otimes \rho ^{AB2}_{ch}\rightarrow \sum _{i}A_{i}\otimes B_{i}(\left| \chi _{j}\right\rangle \left\langle \chi _{j}\right| )^{1}\otimes \rho ^{AB2}_{ch}A_{i}^{\dagger }\otimes B_{i}^{\dagger }\: (j=1,2)
\end{equation}
where \( \sum _{i}(A_{i}^{\dagger }\otimes B_{i}^{\dagger })(A_{i}\otimes B_{i})=I \),
\( A_{i} \)'s being operators on the Hilbert space of Alice's particles and
\( B_{i} \)'s on Bob's. 

Now \( \rho ^{AB2}_{ch} \) may be thought of as traced out from some pure state
\( \left| \Psi \right\rangle ^{AB2M} \) where M is an ancilla \cite{7}. So
now the protocol may be written as 
\begin{equation}
\label{2}
P\left[ \left| \chi _{j}\right\rangle ^{1}\otimes \left| \Psi \right\rangle ^{AB2M}\right] \rightarrow \sum _{i}A_{i}\otimes B_{i}\otimes I^{M}P\left[ \left| \chi _{j}\right\rangle ^{1}\otimes \left| \Psi \right\rangle ^{AB2M}\right] A_{i}^{\dagger }\otimes B_{i}^{\dagger }\otimes I^{M}
\end{equation}
where \( I^{M} \) is the identity operator on the Hilbert space of the ancilla
M. Now for any such protocol, there always exists a unitary operator \( U \)
on the combined Hilbert space of 1, A, B, 2, M and the environment E which executes
the transformation 
\begin{equation}
\label{3}
\left| \chi _{j}\right\rangle ^{1}\otimes \left| \Psi \right\rangle ^{AB2M}\otimes \left| 0\right\rangle ^{E}\rightarrow \sum _{i}\left( A_{i}\otimes B_{i}\otimes I^{M}\left| \chi _{j}\right\rangle ^{1}\otimes \left| \Psi \right\rangle ^{AB2M}\right) \otimes \left| i\right\rangle ^{E}
\end{equation}
so that tracing over the environment, we recover the transformation (2) \cite{8}.
\( \left| 0\right\rangle  \) is a standard state and the \( \left| i\right\rangle  \)'s
are mutually orthonormal states of the environment. 

Since \( \left| \chi _{1}\right\rangle  \) and \( \left| \chi _{2}\right\rangle  \)
are non-orthogonal and as the teleported state which is to surface at particle
2 is pure, we must have 
\[
U\left| \chi _{1}\right\rangle ^{1}\otimes \left| \Psi \right\rangle ^{AB2M}\otimes \left| 0\right\rangle ^{E}=\left| \chi _{1}\right\rangle ^{2}\otimes \left| \Psi ^{\prime }\right\rangle ^{1ABME}\]
\[
U\left| \chi _{2}\right\rangle ^{1}\otimes \left| \Psi \right\rangle ^{AB2M}\otimes \left| 0\right\rangle ^{E}=\left| \chi _{2}\right\rangle ^{2}\otimes \left| \Psi ^{\prime }\right\rangle ^{1ABME}\]
for some state \( \left| \Psi ^{\prime }\right\rangle  \). 

Now if 
\[
\left| \chi \right\rangle =a_{1}\left| \chi _{1}\right\rangle +a_{2}\left| \chi _{2}\right\rangle \]
where \( a_{1} \) and \( a_{2} \) are arbitrary comclex numbers satisfying
\( \left\langle \chi \right. \left| \chi \right\rangle =1 \), then
\begin{equation}
\label{4}
U\left| \chi \right\rangle ^{1}\otimes \left| \Psi \right\rangle ^{AB2M}\otimes \left| 0\right\rangle ^{E}=\left| \chi \right\rangle ^{2}\otimes \left| \Psi ^{\prime }\right\rangle ^{1ABME}
\end{equation}
so that \( \left| \chi \right\rangle  \) is also teleported by the same protocol
and channel which teleports \( \left| \chi _{1}\right\rangle  \) and \( \left| \chi _{2}\right\rangle  \).
But \( \left| \chi \right\rangle  \) is an arbitrary qubit. And teleporting
an arbitrary pure qubit requires maximal entanglement as was discussed in ref.\cite{5},
using a result of ref.\cite{3}. Thus if the state to be teleported is one of
two known non-orthogonal pure qubits, the channel must be maximally entangled
in \( 2\otimes 2 \). 

Next consider the case when Alice has the task to teleport a state which is
known to be one of the two noncommuting qubits \( \rho _{1} \) and \( \rho _{2} \).
As before, any local protocol which Alice and Bob may choose can be represented
as 
\begin{equation}
\label{5}
\rho ^{1}_{j}\otimes \rho ^{AB2}_{ch}\rightarrow \sum _{i}A_{i}\otimes B_{i}\rho ^{1}_{j}\otimes \rho ^{AB2}_{ch}A_{i}^{\dagger }\otimes B_{i}^{\dagger }\: (j=1,2)
\end{equation}
where \( \sum _{i}(A_{i}^{\dagger }\otimes B_{i}^{\dagger })(A_{i}\otimes B_{i})=I \),
\( A_{i} \)'s being operators on the Hilbert space of Alice's particles and
\( B_{i} \)'s on Bob's. 

Now if \( \rho _{1} \) and \( \rho _{2} \) are to be teleported by this protocol,
we must have 
\begin{equation}
\label{6}
Tr_{1AB}\left( \sum _{i}A_{i}\otimes B_{i}\rho ^{1}_{j}\otimes \rho ^{AB2}_{ch}A_{i}^{\dagger }\otimes B_{i}^{\dagger }\right) =\rho ^{2}_{j}\: (j=1,2)
\end{equation}
Since \( \rho _{1} \) and \( \rho _{2} \) are noncommuting qubits, there (uniquely)
exist two non-orthogonal pure states \( \left| \psi \right\rangle  \) and \( \left| \phi \right\rangle  \)
such that 
\begin{equation}
\label{7}
\rho _{j}=\lambda _{j}\left| \psi \right\rangle \left\langle \psi \right| +(1-\lambda _{j})\left| \phi \right\rangle \left\langle \phi \right| \: (j=1,2)
\end{equation}
where \( 0\leq \lambda _{j}\leq 1 \) and \( \lambda _{1}\neq \lambda _{2} \).
We call these (i.e., \( \left| \psi \right\rangle  \) and \( \left| \phi \right\rangle  \))
the ``extreme'' states corresponding to \( \rho _{1} \) and \( \rho _{2} \).\cite{9}
Therefore
\[
\lambda _{j}Tr_{1AB}\left( \sum _{i}A_{i}\otimes B_{i}(\left| \psi \right\rangle \left\langle \psi \right| )^{1}\otimes \rho ^{AB2}_{ch}A_{i}^{\dagger }\otimes B_{i}^{\dagger }\right) \]
 
\[
+(1-\lambda _{j})Tr_{1AB}\left( \sum _{i}A_{i}\otimes B_{i}(\left| \phi \right\rangle \left\langle \phi \right| )^{1}\otimes \rho ^{AB2}_{ch}A_{i}^{\dagger }\otimes B_{i}^{\dagger }\right) \]
\[
=\lambda _{j}(\left| \psi \right\rangle \left\langle \psi \right| )^{2}+(1-\lambda _{j})(\left| \phi \right\rangle \left\langle \phi \right| )^{2}\]
which may be rewritten as 
\[
\lambda _{j}[Tr_{1AB}\left( \sum _{i}A_{i}\otimes B_{i}(\left| \psi \right\rangle \left\langle \psi \right| )^{1}\otimes \rho ^{AB2}_{ch}A_{i}^{\dagger }\otimes B_{i}^{\dagger }\right) -(\left| \psi \right\rangle \left\langle \psi \right| )^{2}\]
\[
-Tr_{1AB}\left( \sum _{i}A_{i}\otimes B_{i}(\left| \phi \right\rangle \left\langle \phi \right| )^{1}\otimes \rho ^{AB2}_{ch}A_{i}^{\dagger }\otimes B_{i}^{\dagger }\right) +(\left| \phi \right\rangle \left\langle \phi \right| )^{2}]\]
\[
+[Tr_{1AB}\left( \sum _{i}A_{i}\otimes B_{i}(\left| \phi \right\rangle \left\langle \phi \right| )^{1}\otimes \rho ^{AB2}_{ch}A_{i}^{\dagger }\otimes B_{i}^{\dagger }\right) -(\left| \phi \right\rangle \left\langle \phi \right| )^{2}]=0\]
Since \( \lambda _{1}\neq \lambda _{2} \), we must have
\[
Tr_{1AB}\left( \sum _{i}A_{i}\otimes B_{i}(\left| \psi \right\rangle \left\langle \psi \right| )^{1}\otimes \rho ^{AB2}_{ch}A_{i}^{\dagger }\otimes B_{i}^{\dagger }\right) =(\left| \psi \right\rangle \left\langle \psi \right| )^{2}\]
and 
\[
Tr_{1AB}\left( \sum _{i}A_{i}\otimes B_{i}(\left| \phi \right\rangle \left\langle \phi \right| )^{1}\otimes \rho ^{AB2}_{ch}A_{i}^{\dagger }\otimes B_{i}^{\dagger }\right) =(\left| \phi \right\rangle \left\langle \phi \right| )^{2}\]
Thus if we assume that two noncommuting qubits are teleported by some protocol
and through some channel \( \rho _{ch} \), then the corresponding ``extreme''
non-orthogonal pure qubits are also teleported by the same protocol and through
the same channel. 

But we have already proved that to teleport non-orthogonal pure qubits, a maximally
entangled channel is necessary. Thus a maximally entangled channel in \( 2\otimes 2 \)
is necessary even if the teleported state is known to be one of two noncommuting
qubits. 

In ref.\cite{5} it was shown that if Alice has the task of teleporting (to
Bob) a qubit (generalized to higher dimensions in ref.\cite{6}) which is \emph{a
priori} known to belong to a commuting set, then a certain classically correlated
channel (between Alice and Bob) is sufficient. It was further shown that if
Alice has to teleport a qubit known to be one of two arbitrary but fixed noncommuting
qubits, then the channel between Alice and Bob must be entangled. We show here
that in the latter case, the amount of entanglement must be 1 ebit. 

U.S. thanks Dipankar Home for encouragement and acknowledges partial support
by the Council of Scientific and Industrial Research, Government of India, New
Delhi.


\begin{thebibliography}{}
\bibitem{1}W. K. Wootters and W. H. Zurek, Nature \textbf{229}, 802 (1982); D. Dieks, Phys.
Lett. \textbf{92A}, 271 (1982); H. P. Yuen, Phys. Lett. \textbf{113A}, 405 (1986)
\bibitem{2}C. H. Bennett, G. Brassard, C. Crepeau, R. Jozsa, A. Peres, and W. K. Wootters,
Phys. Rev. Lett. \textbf{70}, 1895 (1993)
\bibitem{3}H. F. Chau and H.-K. Lo, \emph{How much does it cost to teleport?}, quant-ph/9605025
\bibitem{4}L. Henderson, L. Hardy and V. Vedral, \emph{Two State Teleportation}, quant-ph/9910028
\bibitem{5}S. Ghosh, G. Kar, A. Roy and U. Sen, \emph{Entanglement vs. Noncommutativity
in Teleportation}, quant-ph/0010012
\bibitem{6}S. Ghosh, G. Kar, A. Roy, D. Sarkar and U. Sen, in \emph{}preparation
\bibitem{7}B. Schumacher, Phys. Rev. A, \textbf{54}, 2614 (1996)
\bibitem{8}K. Kraus, \emph{States, Effects, and Operations: Fundamental Notions of Quantum
Theory} (Springer-Verlag, Berlin, 1983); see also J. Preskill, Lecture Notes
ph229 available at http://www.theory.caltech.edu/people/preskill/ph229/ and
the appendix of \cite{7}.
\bibitem{9}A decomposition like that of Eq. (7) is of course possible for \emph{commuting}
qubits \( \rho _{1} \) and \( \rho _{2} \), in which case \( \left| \psi \right\rangle  \)
and \( \left| \phi \right\rangle  \) are orthogonal.
\end{thebibliography}
\end{document}